# Liquid-liquid equilibria of systems containing 2-methoxyphenol or 2-ethoxyphenol and *n*-alkanes

Cristina Alonso Tristán[a], João Victor Alves-Laurentino[b], Fatemeh Pazoki[b], Susana Villa[b], Daniel Lozano-Martín[b], Fernando Hevia*,[b]

[a] Departamento de Ingeniería Electromecánica, Escuela Politécnica Superior, Universidad de Burgos. Avda. Cantabria s/n, 09006 Burgos, Spain.

[b] GETEF. Departamento de Física Aplicada, Facultad de Ciencias, Universidad de Valladolid. Paseo de Belén, 7, 47011 Valladolid, Spain.

* Corresponding author, e-mail: luisfernando.hevia@uva.es

## Abstract

Liquid-liquid equilibria phase diagrams have been determined for the systems: 2-methoxyphenol + *n*-decane, or + *n*-dodecane, or + *n*-tetradecane or + *n*-hexadecane and for 2-ethoxyphenol + *n*-octane, or + *n*-dodecane, or + *n*-tetradecane, or + *n*-hexadecane. The experimental method used is based on the observation, by means of a laser scattering technique, of the turbidity produced on cooling when a second phase appears. All the mixtures studied show an upper critical solution temperature, which increases with the *n*-alkane size. Dipolar interactions between like molecules become stronger in the sequence: 2-ethoxyphenol < 2-methoxyphenol < phenol. Data available in the literature suggest that this relative variation is also valid for *n*-alkane mixtures containing other substituted anilines, characterized by having a second polar group. The dependence of the upper critical solution temperature on the molecular structure of the polar aromatic compound involved is shortly discussed in terms of intramolecular and steric effects.

# 1. Introduction

Phenol derivatives are crucial intermediates to synthesize a wide variety of specialty chemicals. Phase behaviour of those product mixtures plays an important role in development of the manufacturing processes. Alkoxyphenols have a variety of applications. 2-methoxyphenol is used as an intermediate in the chemical synthesis of active pharmaceuticals, flavouring and perfumery products. For example, 2-methoxyphenol acts as an expectorant [1,2]. It serves as a precursor to flavourings such as eugenol and vanillin [3,4]. It is used as an antioxidant as well [5]. In addition to this, it is also used as an indicator in chemical reactions that produce oxygen [6]. 2-ethoxyphenol is used as an intermediate in chemical synthesis, mainly in pharmaceutical and food industries. Its derivate ethylvanillin is used for fragrance and flavouring among others as a flavourer in perfumes and in the production of chocolate [7]. Phenolic compounds are chemical pollutants that can be found in water, often due to industrial and agricultural wastewater discharge [8]. Thus, their presence in water can be a concern for public health. In this context, it is important to remove these compounds from water. A possibility is liquid-liquid extraction using 1-alkanols, such 1-pentanol or 1-octanol [8]. The corresponding design of the process needs a deep understanding of the interactions between the groups involved. At this end, the experimental characterization of phenolic compound + $n$-alkane is the first step for a further theoretical study of mixtures including phenolic compounds.

For several years, our group has been interested on the investigation of intramolecular effects between the phenyl ring ($C_6H_5$-group) and a polar group, X, more or less directly attached to an aromatic ring, as these effects are very different from those between the same groups when they belong to different molecules. For example, the upper critical solution temperature (UCST) of the system (phenol + $n$-decane) is 336.50 K [9], while the excess molar enthalpy, $H_m^E$, of the system (decan-1-ol + benzene) at equimolar composition and temperature $T$ = 298.15 K is 1050 J·mol$^{-1}$ (extrapolated value from the literature [10]). That is, intramolecular effects lead to enhanced interactions between like polar molecules, while intermolecular effects are usually more favourable in interactions between unlike molecules. For this purpose, we have studied intramolecular proximity effects in mixtures containing aromatic amines (anilines [11,12], 2-amino-1-methylbenzene [13], 1-phenylmethanamine [14,15], 2-ethoxy-benzenamine [16], 4-ethoxy-benzenamine [16], 1H-pyrrole [17], quinoline [18] or imidazoles [19]); aldehydes and ketones [20–23] (phenyl methanal, 1-phenylethanone, 4-phenyl-2-butanone), esters [23] (benzyl ethanoate), nitriles [24] (benzonitrile, phenyl acetonitrile, 3-phenylpropionitrile), 2-phenoxy-ethanol [25] or aromatic alkanols [26–28] (phenol, phenylmethanol, 2-phenylethan-1-ol).

Alkoxyphenol molecules contain two different polar groups (-OH and -O-) attached to the phenyl ring in different positions, and intramolecular proximity effects between these two groups are also expected to exist. The case of (phenoxyethanol + $n$-alkane) systems, examined in a previous work [25], is somewhat different, as the -O- and -OH groups are placed in the same linear chain attached to the phenyl ring. As a continuation, we provide now liquid-liquid equilibrium (LLE) data for $n$-alkane systems involving 2-methoxyphenol (guaiacol) or 2-ethoxyphenol (guaethol). More precisely, we report, at pressure $p$ = 0.1 MPa, the equilibrium composition and temperature of the systems: 2-methoxyphenol + $n$-decane, or + $n$-dodecane, or + $n$-tetradecane or + $n$-hexadecane; and 2-ethoxyphenol + $n$-octane, or + $n$-dodecane, or + $n$-tetradecane, or + $n$-hexadecane.

# 2. Materials and methods

## 2.1. Materials

All the information related to the source, purity, water content and density, $\rho$, of the pure compounds used in the present experimental research is collected in Table 1. The chemicals were used as received. The measured $\rho$ values, listed in Table 1, show that our results are in good agreement with values from the literature. After a careful survey of literature data, no previous $\rho$ value has been found for 2-ethoxyphenol. 2-methoxyphenol presents subcooling effect at room temperature and its purity was tested also by measuring its melting temperature. The result obtained was 301.4 K, agreeing with literature values (from 301.15 K to 301.51 K [29]).

Water mass fractions, $w_{H_2O}$, of the pure liquids were determined by the Karl-Fischer method with a relative standard uncertainty of 0.02. Density measurements were conducted by means of a standard vibrating-tube densimeter model Anton Paar DSA 5000 with a temperature stability of 0.001 K. The standard uncertainties of temperature and density are 0.01 K and 0.10 kg·m$^{-3}$ respectively. A modulated differential scanning calorimeter, model TA Instruments Q2000, was used to measure the melting temperature of 2-methoxyphenol. Tzero aluminium pans and lids purchased from TA Instruments were filled with ∼ 7 mg of sample, and hermetically sealed. The seal prevented loss of mass from volatilization of the sample or operator error. The estimated standard uncertainty of the melting temperature is 0.3 K.

**Table 1**

Properties of pure compounds at pressure $p$ = 0.1 MPa and temperature $T$ = 298.15 K: initial purity, density, $\rho$, and water mass fraction, $w_{H_2O}$. [a]

| Compound | CAS | Source | Initial purity[b] | $\rho$ /kg·m$^{-3}$ Exp. | $\rho$ /kg·m$^{-3}$ Lit. | $10^6 \cdot w_{H_2O}$ |
|---|---|---|---|---|---|---|
| 2-methoxyphenol | 90-05-1 | Sigma-Aldrich | ≥ 0.98 | 1128.50 | 1128.9 [29] | 64 |
|  |  |  |  |  | 1129 [30] |  |
| 2-ethoxyphenol | 94-71-3 | Sigma-Aldrich | ≥ 0.98 | 995.93 |  | 32 |
| $n$-octane | 111-65-9 | Sigma-Aldrich | ≥ 0.99 | 698.68 | 698.62 [31] | 38 |
| $n$-decane | 124-18-5 | Fluka | ≥ 0.99 | 726.35 | 726.35 [31] | 20 |
| $n$-dodecane | 112-40-3 | Fluka | ≥ 0.98 | 745.51 | 745.31 [32] | 25 |
| $n$-tetradecane | 629-59-4 | Fluka | ≥ 0.99 | 759.27 | 759.29 [32] | 25 |
| $n$-hexadecane | 544-76-3 | Fluka | ≥ 0.99 | 770.22 | 770.06 [32] | 33 |
|  |  |  |  |  | 770.2 [33] |  |
|  |  |  |  |  | 770.79 [34] |  |

[a] Expanded uncertainties ($U$) with a coverage factor of 2: $U(T)$ = 0.02 K; $U(p)$ = 10 kPa; $U(\rho)$ = 0.20 kg·m$^{-3}$; $U(w_{H_2O})$ = 0.04 · $w_{H_2O}$.
[b] Gas chromatography area fraction. Provided by the supplier.

## 2.2. Apparatus and procedure for LLE measurements

Small Pyrex tubes (0.009 m of diameter and about 0.04 m of length; free volume of the ampoule ≈ 1.17 · 10$^{-6}$ m$^3$) were used for the preparation of the mixtures. The composition of the mixtures, given as a mole fraction of the alkoxyphenol (2-methoxyhenol or 2-ethoxyphenol), $x_1$, was calculated from mass measurements, performed with an analytical balance Sartorius MSU125P and correcting for buoyancy effects, with a standard uncertainty of 5·10$^{-8}$ kg. The mentioned tubes are immediately sealed by capping at pressure $p$ = 0.1 MPa and temperature $T$ = 298.15 K, ready to use in the LLE measurements.

Temperatures of the samples during the experiments were measured with a Pt-1000 resistance calibrated using the triple point of water and the melting point of Ga as reference standards (standard uncertainty of 0.001 K).

The LLE curves were determined by means of the observation of the turbidity produced on cooling when a second phase appears. A summary of the procedure follows. (i) The samples in the sealed Pyrex tubes are placed in a thermostatic bath few tenths of K above the expected temperature. (ii) Mixtures are then slowly cooled at a rate of 1.2 K·h$^{-1}$ under continuous stirring. Since the equilibrium times are much longer in the two-phase region than the corresponding times in the one-phase region, this method is suitable to prevent subcooling and gravity effects in mixtures at compositions far from the critical one [35,36]. (iii) A red He-Ne laser is situated on one side of the equilibrium cell, and the light beam crossing through the solution is focused on a photodiode placed at the other side of the cell (see below). When the temperature is slowly decreased, small drops of the dispersed liquid phase start to grow, and the light is dispersed during the transition. This causes a voltage variation in the mentioned photodiode, which is determined by a digital Agilent 34410A multimeter connected to a PC. Transition temperatures can be then measured. (iv) Two or three runs are usually conducted to get a better assessment of the equilibrium temperatures.

The measuring setup was tested by measuring the test system (methanol + heptane), obtaining deviations of the critical mole fraction lower than 0.001 and of the critical temperature lower than 0.2 K, according to reference data from literature. A direct comparison between results obtained using our experimental technique and data from the literature can be encountered elsewhere [37]. The laser used in this technique is a self-contained Helium - Neon laser from Uniphase, model Novette 1508-0. Its key specifications are a minimum output power of 0.5 mW, wavelength of 632.8 nm, beam diameter of 0.48 mm, beam divergence of 1.7 mrad and mode (TEM$_{00}$) purity greater than 95 %. The photodiode used is an integrated photodiode and transimpedance amplifier from Texas Instruments, model OPT301M. It is packaged in a hermetic TO-99 metal package with a glass window. Its key specifications are a responsivity of 0.47 A/W at 650 nm, dark errors of 2 mV, bandwidth of 4 kHz and quiescent current of 400 μA.

The expanded uncertainties $U$ (with a coverage factor of 2) related to the LLE measurements are the following. The uncertainty of pressure is $U(p)$ = 10 kPa. For the equilibrium mole fractions, the estimated uncertainty is $U(x_1)$ = 0.0010; this value already takes into consideration the partial evaporation in the mentioned free volume of the ampoule. The estimated uncertainty of the equilibrium temperature depends on the region where measurements are conducted. In the flat region of the coexistence curves (top of the curves), the uncertainty of the temperature is $U(T)$ = 0.20 K, while outside of this region (tails of the curves), it is $U(T)$ = 0.40 K.

## 3. Results

The directly measured LLE temperatures, $T$, and mole fractions of the alkoxyphenol, $x_1$, at pressure $p$ = 0.1 MPa of the studied systems are collected in Table 2 and represented in Figs. 1 and 2. The studied systems are: 2-methoxyphenol + *n*-decane, or + *n*-dodecane, or + *n*-tetradecane or + *n*-hexadecane; and 2-ethoxyphenol + *n*-octane, or + *n*-dodecane, or + *n*-tetradecane, or + *n*-hexadecane. LLE data for comparison are not available in the literature, except for the system (2-methoxyphenol + *n*-hexadecane) [38], which are in good agreement with our results.

**Table 2**

Liquid-liquid equilibrium temperature, $T$, and mole fraction of the alkoxyphenol, $x_1$, of (alkoxyphenol (1) + $n$-alkane (2)) systems at pressure $p$ = 0.1 MPa. [a]

| $x_1$ | $T/K$ | $x_1$ | $T/K$ | $x_1$ | $T/K$ | $x_1$ | $T/K$ |
|---|---|---|---|---|---|---|---|
| 2-methoxyphenol (1) + $n$-decane (2) | | 2-methoxyphenol (1) + $n$-dodecane (2) | | 2-methoxyphenol (1) + $n$-tetradecane (2) | | 2-methoxyphenol (1) + $n$-hexadecane (2) | |
| 0.2909 | 311.76 | 0.3206 | 317.84 | 0.2953 | 315.26 | 0.3875 | 326.65 |
| 0.3179 | 314.07 | 0.3467 | 319.76 | 0.3498 | 320.46 | 0.4021 | 327.73 |
| 0.3325 | 315.17 | 0.3805 | 321.89 | 0.3849 | 324.30 | 0.4234 | 329.48 |
| 0.3564 | 316.70 | 0.4200 | 324.23 | 0.4125 | 326.33 | 0.4487 | 331.23 |
| 0.3775 | 317.81 | 0.4440 | 325.27 | 0.4413 | 328.24 | 0.4655 | 332.43 |
| 0.4008 | 319.13 | 0.4776 | 326.70 | 0.4551 | 329.10 | 0.4973 | 334.33 |
| 0.4410 | 320.53 | 0.4964 | 327.13 | 0.4850 | 330.57 | 0.5368 | 336.56 |
| 0.4621 | 321.25 | 0.5140 | 327.68 | 0.5086 | 331.63 | 0.5632 | 337.68 |
| 0.4811 | 321.63 | 0.5287 | 328.06 | 0.5455 | 333.32 | 0.5981 | 338.99 |
| 0.4954 | 321.73 | 0.5395 | 328.13 | 0.5596 | 333.49 | 0.6273 | 339.72 |
| 0.5055 | 321.92 | 0.5770 | 328.37 | 0.5940 | 334.49 | 0.6574 | 340.50 |
| 0.5234 | 322.31 | 0.6033 | 328.55 | 0.6049 | 334.88 | 0.6807 | 340.60 |
| 0.5451 | 322.32 | 0.6211 | 328.81 | 0.6271 | 334.81 | 0.7138 | 340.82 |
| 0.5766 | 322.66 | 0.6407 | 328.62 | 0.6511 | 334.99 | 0.7353 | 340.92 |
| 0.5784 | 322.72 | 0.6674 | 328.76 | 0.6542 | 335.17 | 0.7534 | 340.95 |
| 0.6071 | 322.72 | 0.6889 | 328.64 | 0.6653 | 335.40 | 0.7753 | 341.03 |
| 0.6197 | 322.78 | 0.7225 | 328.53 | 0.6789 | 335.18 | 0.7881 | 341.00 |
| 0.6342 | 322.71 | 0.7355 | 328.37 | 0.7100 | 334.97 | 0.8127 | 341.00 |
| 0.6691 | 322.54 | 0.7513 | 328.11 | 0.7111 | 334.91 | 0.8252 | 340.88 |
| 0.6940 | 322.30 | 0.7523 | 328.15 | 0.7138 | 334.96 | 0.8469 | 340.17 |
| 0.7135 | 322.26 | 0.7856 | 327.07 | 0.7163 | 334.92 | 0.8638 | 339.18 |
| 0.7311 | 321.95 | 0.8061 | 325.79 | 0.7404 | 335.02 | 0.8787 | 337.80 |
| 0.7512 | 321.31 | 0.8231 | 324.69 | 0.7468 | 334.99 | 0.8963 | 335.43 |
| 0.7767 | 320.32 | 0.8437 | 323.06 | 0.7647 | 334.84 | 0.9138 | 331.55 |
| 0.8023 | 318.61 | 0.8634 | 320.66 | 0.7892 | 334.66 | 0.9288 | 326.54 |
| 0.8258 | 316.16 | | | 0.8131 | 334.60 | 0.9469 | 317.80 |
| 0.8547 | 311.86 | | | 0.8422 | 333.54 | | |
| | | | | 0.8532 | 332.69 | | |
| | | | | 0.8690 | 331.43 | | |
| | | | | 0.8837 | 328.93 | | |
| | | | | 0.9137 | 323.89 | | |
| | | | | 0.9345 | 317.98 | | |
| 2-ethoxyphenol (1) + $n$-octane (2) | | 2-ethoxyphenol (1) + $n$-dodecane (2) | | 2-ethoxyphenol (1) + $n$-tetradecane (2) | | 2-ethoxyphenol (1) + $n$-hexadecane (2) | |
| 0.2653 | 275.49 | 0.2385 | 278.90 | 0.3011 | 286.37 | 0.2931 | 288.02 |
| 0.2931 | 276.82 | 0.2791 | 282.10 | 0.3300 | 288.54 | 0.3314 | 290.96 |
| 0.3224 | 277.70 | 0.3150 | 284.51 | 0.3518 | 289.97 | 0.3873 | 294.86 |
| 0.3506 | 278.29 | 0.3352 | 285.84 | 0.3713 | 291.14 | 0.4374 | 297.82 |
| 0.3789 | 278.69 | 0.3580 | 286.99 | 0.3904 | 292.25 | 0.4855 | 299.80 |
| 0.4094 | 279.00 | 0.3925 | 288.44 | 0.4132 | 293.34 | 0.5122 | 300.75 |
| 0.4446 | 279.15 | 0.4351 | 289.70 | 0.4395 | 294.55 | 0.5426 | 301.57 |
| 0.4701 | 279.18 | 0.4631 | 290.37 | 0.4811 | 295.78 | 0.5842 | 302.33 |
| 0.4963 | 279.27 | 0.4825 | 290.75 | 0.5166 | 296.46 | 0.6147 | 302.75 |
| 0.5301 | 279.21 | 0.5107 | 291.06 | 0.5517 | 296.97 | 0.6441 | 302.88 |
| 0.5560 | 279.18 | 0.5483 | 291.31 | 0.5640 | 297.00 | 0.6826 | 303.02 |
| 0.5872 | 278.98 | 0.5778 | 291.36 | 0.5851 | 297.36 | 0.6965 | 303.01 |
| 0.6168 | 278.83 | 0.6045 | 291.39 | 0.5952 | 297.37 | 0.7218 | 302.95 |
| 0.6180 | 278.81 | 0.6483 | 291.39 | 0.6208 | 297.55 | 0.7468 | 302.94 |
| 0.6415 | 278.49 | 0.6757 | 291.31 | 0.6456 | 297.46 | 0.7700 | 302.86 |
| 0.6423 | 278.53 | 0.6967 | 291.29 | 0.6707 | 297.56 | 0.7958 | 302.80 |
| 0.6750 | 277.75 | 0.7254 | 291.14 | 0.7020 | 297.45 | 0.8220 | 302.54 |
| 0.7046 | 276.88 | 0.7538 | 290.68 | 0.7291 | 297.35 | 0.8478 | 301.92 |
| 0.7368 | 275.57 | 0.7734 | 290.23 | 0.7537 | 297.15 | 0.8663 | 300.86 |
| | | 0.8089 | 288.86 | 0.7751 | 296.88 | 0.8867 | 299.05 |
| | | 0.8267 | 287.63 | 0.8055 | 296.26 | 0.8986 | 297.30 |
| | | 0.8485 | 285.65 | 0.8324 | 295.06 | 0.9027 | 296.57 |
| | | 0.8562 | 284.68 | 0.8499 | 294.11 | 0.9182 | 293.00 |
| | | 0.8657 | 283.53 | 0.8692 | 292.45 | 0.9363 | 286.55 |
| | | 0.8690 | 282.98 | 0.8809 | 291.12 | | |
| | | | | 0.8950 | 289.14 | | |
| | | | | 0.9064 | 287.21 | | |

[a] Expanded uncertainties ($U$) with a coverage factor of 2: $U(p)$ = 10 kPa; $U(x_1)$ = 0.0010; $U(T)$ = 0.20 K in the flat region of the coexistence curves; $U(T)$ = 0.40 K outside the flat region.

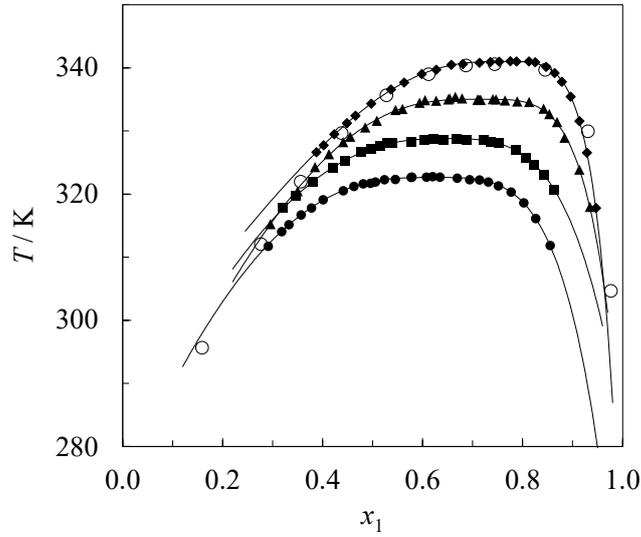

**Fig. 1**. Liquid-liquid equilibrium curves of (2-methoxyphenol + *n*-alkane) systems at pressure $p$ = 0.1 MPa. Symbols, experimental data: (●) *n*-decane (this work); (■) *n*-dodecane (this work); (▲) *n*-tetradecane (this work); (♦) *n*-hexadecane (this work); (o) *n*-hexadecane [38]. Solid lines, calculations with Eq. (1) using the parameters listed in Table 3.

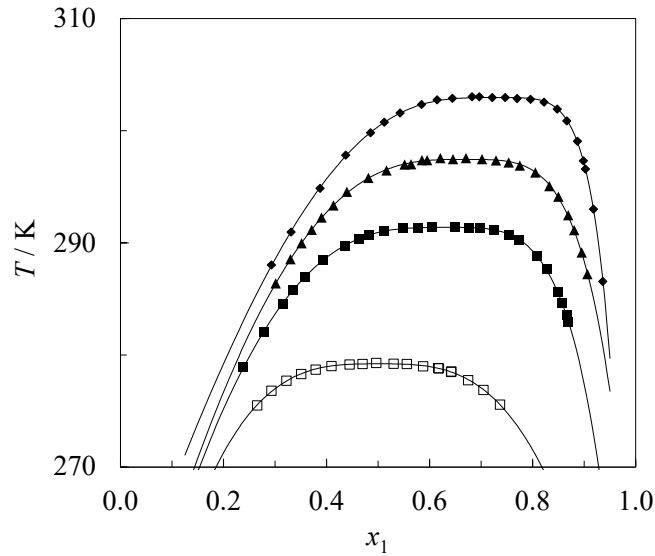

**Fig. 2**. Liquid-liquid equilibrium curves of (2-ethoxyphenol + *n*-alkane) systems at pressure $p$ = 0.1 MPa. Full symbols, experimental data (this work): (□) *n*-octane; (■) *n*-dodecane; (▲) *n*-tetradecane; (♦) *n*-hexadecane. Solid lines, calculations with Eq. (1) using the parameters listed in Table 3.

The experimental ($x_1$, $T$) data of each system were correlated by means of the equation [39,40]:

$$T/\text{K} = T_c/\text{K} + k|y - y_c|^m \tag{1}$$

where

$$y = \frac{\alpha\, x_1}{1 + x_1(\alpha - 1)} \tag{2}$$

$$y_c = \frac{\alpha\, x_{1c}}{1 + x_{1c}(\alpha - 1)} \tag{3}$$

and ($x_{1c}$, $T_c$) represent the composition and temperature of the system at the critical point. More details about subtleties of Eqs. (1) to (3) and the meaning of the parameters are included in a previous work [37]. When $\alpha$ = 1, Eq. (1) is similar to the well-known equation [41–43]:

$$\Delta\lambda_1 = B\tau^\beta \tag{4}$$

In Eq. (4), $\tau = (T_c - T)/T_c$ is the reduced temperature, $\Delta\lambda_1 = \lambda_1' - \lambda_1''$ is the so-called order parameter (which can be the difference between the values $\lambda_1'$ and $\lambda_1''$ in the two phases of any density variable $\lambda_1$ [42]; in our case, $\lambda_1 = x_1$) and $\beta$ is a critical exponent corresponding to this order parameter (whose value depends on the theory applied to its determination [37,42]).

In Eqs. (1) to (3), the parameters $m$, $k$, $\alpha$, $T_c$ and $x_{1c}$ are fitted to the experimental data by a Marquadt algorithm [44] with all the points weighed equally. The root-mean-square deviation of the fit to equation (1), $\sigma(T)$, is defined by:

$$\sigma(T) = \left[\frac{1}{N-n}\sum_{i=1}^{N}\left(T_i^{\text{calc}} - T_i^{\text{exp}}\right)^2\right]^{1/2} \tag{5}$$

where $N$ and $n$ stand for the number of data points and the number of fitted parameters, respectively; $T_i^{\text{calc}}$ refers to the calculated values from Eq. (1) and $T_i^{\text{exp}}$ denotes the experimental values. The values of the parameters and $\sigma(T)$ are listed in Table 3.

**Table 3**

Parameters ($m$, $k$, $\alpha$, $T_c$, $x_{1c}$) of Eq. (1) and root-mean-square deviation ($\sigma(T)$, Eq. (5)) of the fit of the liquid-liquid equilibrium temperature ($T$) as a function of the alkoxyphenol mole fraction ($x_1$), for 2-methoxyphenol, or 2-ethoxyphenol (1) + $n$-alkane (2) mixtures at pressure $p$ = 0.1 MPa. $N$ is the number of experimental points used for the fit, $T_c$ is the upper critical solution temperature and $x_{1c}$ is the critical composition.

| System | N | m | k | $\alpha$ | $T_c$/K | $x_{1c}$ | $\sigma(T)$/K |
|---|---|---|---|---|---|---|---|
| 2-methoxyphenol (1) + $n$-decane (2) | 27 | 3.2225 | -585.32 | 0.5182 | 322.63 | 0.6252 | 0.07 |
| 2-methoxyphenol (1) + $n$-dodecane (2) | 25 | 2.9390 | -377.98 | 0.5389 | 328.75 | 0.6523 | 0.13 |
| 2-methoxyphenol (1) + $n$-tetradecane (2) | 32 | 3.3945 | -627.75 | 0.4050 | 335.02 | 0.7169 | 0.13 |
| 2-methoxyphenol (1) + $n$-hexadecane (2) | 26 | 3.3648 | -692.23 | 0.2676 | 340.99 | 0.7614 | 0.06 |
| 2-ethoxyphenol (1) + $n$-octane (2) | 19 | 3.1438 | -343.54 | 1.0112 | 279.20 | 0.5006 | 0.04 |
| 2-ethoxyphenol (1) + $n$-dodecane (2) | 25 | 3.4908 | -557.07 | 0.5428 | 291.37 | 0.6320 | 0.04 |
| 2-ethoxyphenol (1) + $n$-tetradecane (2) | 27 | 3.3927 | -476.88 | 0.5091 | 297.43 | 0.6713 | 0.06 |
| 2-ethoxyphenol (1) + $n$-hexadecane (2) | 24 | 3.9204 | -917.07 | 0.3246 | 302.95 | 0.7311 | 0.05 |

# 4. Discussion

In this section, the values of the thermophysical properties will be considered at $T$ = 298.15 K and $x_1 = 0.5$ unless otherwise specified. We will denote by $n$ the number of carbon atoms of the $n$-alkane.

As in many systems previously investigated [5,6,8,12–17,19,20] the LLE curves of the mixtures under study are characterized by some typical features: (i) they show a flat maximum (Figs. 1 and 2); (ii) for a given alkoxyphenol, the curves become progressively shifted towards higher $x_1$ values when $n$ increases (Figs. 1 and 2); (iii) for a given alkoxyphenol, the upper critical solution temperature (UCST) increases roughly linearly with $n$ (Table 3, Figs. 1-3). Feature (iii) is observed in mixtures formed by $n$-alkane and one of the following types of compounds: linear alkanone [45], linear organic carbonate [46], acetic anhydride [47], alkoxyethanol [37,48,49], polyether [50,51], amide [52,53], aniline [54,55], azapan-2-one [56], phenol [9,55,57,58], aromatic alcohols [27,28], phenoxyalcohol [25], benzaldehyde [20], aromatic alkanone [21–23], 2-hydroxy benzaldehyde [59] and phenetidines [16].

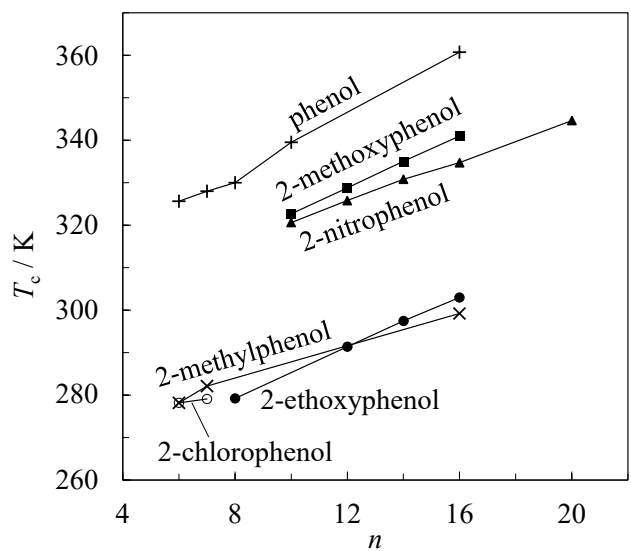

**Fig. 3**. Upper critical solution temperatures ($T_c$) of (aromatic polar compound + $n$-alkane) systems at pressure $p$ = 0.1 MPa as a function of the number of carbon atoms of the $n$-alkane ($n$): (+) phenol [9,38,55,58]; (■) 2-methoxyphenol (this work, [38]); (▲) 2-nitrophenol [38,60]; (●) 2-ethoxyphenol (this work); (x) 2-methylphenol [60,61]; (○) 2-chlorophenol [60]. See also Table 4.

For a fixed $n$-alkane, we note that UCST changes in the order: 2-methoxyphenol > 2-ethoxyphenol. We have shown along a series of investigations that, in $n$-alkane systems, dipolar interactions between aromatic polar molecules involving a polar group X are stronger than those between isomeric linear molecules with the same X group. For example, UCST(phenol + $n$-heptane) = 327.3 K [58], while (hexan-1-ol + $n$-heptane) show no miscibility gap at $T$ = 298.15 K and $H_m^E$(hexan-1-ol + $n$-heptane) = 527 J mol$^{-1}$ [62]. This behaviour has been ascribed to the existence of intramolecular proximity effects between the $C_6H_5$- group and the polar X group under consideration.

When the aromatic ring and the polar group X are not in the same molecule, i.e., in a mixture of the form (polar compound + aromatic compound), the intermolecular effects that appear between the polar compound and the aromatic compound can be relevant compared to the corresponding (polar compound + *n*-alkane) systems. In fact, the $H_\mathrm{m}^\mathrm{E}$ values of (alkan-1-ol + benzene) systems are higher than those of the corresponding *n*-alkane mixtures. For example, $H_\mathrm{m}^\mathrm{E}$(hexan-1-ol + benzene) = 1141 J mol$^{-1}$ [63] > $H_\mathrm{m}^\mathrm{E}$(hexan-1-ol + *n*-hexane) = 461 J mol$^{-1}$ [64]. From these values, it is clear that benzene is a much more effective breaker of the alcohol self-association than *n*-alkanes. However, when the aromatic ring and the polar group are in the same molecule, the intramolecular proximity effects can lead to enhanced intermolecular interactions between like molecules. This is the case of the (alkoxyphenol + *n*-alkane) solutions, where there are 3 coupled functional groups in the same molecule (aromatic ring, -OH and methoxy or ethoxy group). The significantly enhanced alkoxyethanol-alkoxyethanol interactions lead to miscibility gaps, as demonstrated by our LLE data.

In mixtures with a given *n*-alkane, UCSTs decrease when the aromatic polar compound (phenol, X = OH) is replaced by the corresponding 2-methyl, 2-cloro, 2-nitro, 2-methoxy or 2-ethoxy derivative (2-methylphenol, 2-clorophenol, 2-nitrophenol, 2-methoxyphenol or 2-ethoxyphenol) (Table 4, Fig. 3). This behaviour can be ascribed to the lower aromatic surface fraction of the phenol derivatives compared to phenol, which leads to weaker proximity effects between the -OH group and the phenyl ring.

In the analogous case of solutions containing 2-ethoxy-benzenamine or 4-ethoxy-benzenamine and *n*-alkanes [16], it is observed that the UCSTs of 2-ethoxy-benzenamine mixtures are lower than that of the systems with 4-ethoxy-benzenamine, due to the proximity between the amine group and the ethoxyl group in 2-ethoxy-benzenamine, which hinders the amine group and weakens interactions between like molecules. This suggests that a similar steric effect should be present in (alkoxyphenol + *n*-alkane) mixtures.

**Table 4**

Upper critical solution temperatures ($T_\mathrm{c}$) for (aromatic polar compound + *n*-alkane) mixtures. The polar compounds considered are phenol (C$_6$H$_5$-X, polar group X = OH), 2-methylphenol and its derivatives including the Y group in the 2-position (2-Y-C$_6$H$_4$-X, for Y = CH$_3$, Cl, NO$_2$, OCH$_3$, OCH$_2$CH$_3$). Non-referenced values are experimental measurements from the present work.

| *n*-alkane | $T_\mathrm{c}$(C$_6$H$_5$-X + *n*-alkane)/K | Y | 2-Y-C$_6$H$_4$-X | $T_\mathrm{c}$(2-Y-C$_6$H$_4$-X + *n*-alkane)/K |
|---|---|---|---|---|
| *n*-hexane | 325.6 [55] | CH$_3$ | 2-methylphenol | 278.1 [60] |
| | | Cl | 2-clorophenol | 278.1 [60] |
| *n*-heptane | 327.3 [58] | CH$_3$ | 2-methylphenol | 282.1 [60] |
| | | Cl | 2-clorophenol | 279.1 [60] |
| | | NO$_2$ | 2-nitrophenol | 316.2 [60] |
| *n*-octane | 329.5 [58] | OCH$_2$CH$_3$ | 2-ethoxyphenol | 279.2 |
| *n*-decane | 336.5 [9] | OCH$_3$ | 2-methoxyphenol | 322.6 |
| *n*-hexadecane | 360.7 [38] | CH$_3$ | 2-methylphenol | 299.2 [61] |
| | | OCH$_3$ | 2-methoxyphenol | 340.5 [38], 341.0 |
| | | OCH$_2$CH$_3$ | 2-ethoxyphenol | 302.9 |
| | | NO$_2$ | 2-nitrophenol | 334.6 [38] |

Let us examine the effect on the UCST of the replacement of the nonpolar CH$_3$ group in a 2-methylphenol molecule by a polar group Y = Cl, NO$_2$, OCH$_3$, OCH$_2$CH$_3$, (having then two polar groups in the molecule), in *n*-alkane systems. For that purpose, let us compare the UCSTs of *n*- alkane systems containing 2-methylphenol with the corresponding systems with 2-clorophenol, 2-nitrophenol, 2-methoxyphenol or 2-ethoxyphenol (Table 4, Fig. 3). For *n*-hexadecane systems, UCST values change in the order: 2-methylphenol < 2-ethoxyphenol < 2-nitrophenol < 2-methoxyphenol. In contrast, for *n*-hexane and *n*-heptane systems, the UCSTs vary as: 2-clorophenol ≤ 2-methylphenol. Consequently, dipolar interactions among polar molecules of the same species become stronger in the sequence: 2-clorophenol < 2-methylphenol < 2-ethoxyphenol < 2-nitrophenol < 2-methoxyphenol. This variation can be ascribed to the existence of new intramolecular interactions between the (X, Y) groups, and between the (Y, $C_6H_5$-) groups. This proximity effect seems to reduce the strength of the dipolar interactions in 2-clorophenol compared to 2-methylphenol, whereas it enhances them in 2-ethoxyphenol, 2-nitrophenol and 2-methoxyphenol.

An interesting point is that the UCST of *n*-hexadecane systems changes in the order: 2-ethoxyphenol < 2-methoxyphenol < phenol. In other words, dipolar interactions between the polar molecules are smaller when the new Y group in the 2-position gets larger. This suggests that steric hindrance due to the size of the new Y group in the 2-position are predominant over the X-Y proximity effects.

Also, from this study it can be seen that the most stable conformations for these systems are mainly determined by the balance of two antagonistic phenomena: (i) electron conjugation effects involving the aromatic ring and the π-electron donor methoxyl or ethoxyl group (these interactions are favoured by the coplanarity between the substituents and the ring), and (ii) steric repulsions between ortho groups which force the methoxyl or ethoxyl group to adopt an out of plane conformation relative to the ring. The H atom of the hydroxyl group points toward the ortho methoxyl substituent, a preference that results in further stabilization, since an intramolecular hydrogen bond interaction presumably comes into play with the oxygen on the methoxyl group, the acceptor for the hydroxyl hydrogen donor. For 2-methoxyphenol, the H···O($CH_3$) distance is found to be 2.086 Å [65]. This value, which is lower than the sum of the Van der Waals radii of oxygen (1.4 Å) and hydrogen (1.2 Å), supports the presence of an intramolecular hydrogen bond in 2-methoxyphenol. Also, other experimental (nuclear magnetic resonance [66,67], gas electron diffraction [68], core X-ray photoelectron spectroscopy [69], infrared spectroscopy [67,70,71], thermochemical [71]) and quantum-chemical calculation [68,69,71] studies further support it and compare the effect with the case of 3-methoxyphenol and 4-methoxyphenol [69–71]. Unfortunately, to the best of our knowledge no similar information is available in the literature for phenols with more than one methoxyl substituent, particularly for 2-ethoxyethanol. When an additional methylene group is added to the methoxyl group of 2-methoxyphenol to become 2-ethoxyphenol, the final methyl group of 2-ethoxyphenol should be in gauche position with respect to the benzene ring, however according to two-dimensional Infrared Vibrational Echo Spectroscopy [72], the methyl group is below the plane of benzene ring, i.e., the methyl group is in the anti-position relative to the benzene ring. This addition of a methylene group to the methoxyl group, i.e. to form 2-ethoxyphenol, hinders the formation of the intramolecular hydrogen bond, causing the UCST to decrease.

## 5. Conclusions

LLE phase diagrams have been obtained for systems 2-methoxyphenol + *n*-decane, or + *n*-dodecane, or + *n*-tetradecane or + *n*-hexadecane; and 2-ethoxyphenol + *n*-octane, or + *n*-dodecane, or + *n*-tetradecane, or + *n*-hexadecane. The liquid-liquid equilibrium temperature data have been correlated to Eq. (1) as a function of the alkoxyphenol mole fraction, obtaining root-mean-square deviations of less than 0.1 K for all the mixtures studied.

In this work, intramolecular effects in alkoxyphenol systems have been analyzed and compared with other phenol derivatives. For a given alkoxyphenol, the mixtures are characterized by having UCSTs which increase with the *n*-alkane size. For a given *n*-alkane, the UCST is higher for mixtures containing 2-methoxyphenol. Dipolar interactions between like molecules become stronger in the sequence: 2-ethoxyphenol < 2-methoxyphenol < phenol. The attachment of a second polar group in the ortho position to a phenol leads to weaker dipolar interactions.

# CRediT author statement

**Cristina Alonso Tristán**: Data curation, Formal analysis, Investigation, Writing – original draft. **João Victor Alves-Laurentino**: Investigation, Writing – review & editing. **Fatemeh Pazoki**: Investigation, Writing – review & editing. **Susana Villa**: Formal analysis, Validation, Writing – review & editing. **Daniel Lozano-Martín**: Funding acquisition, Supervision, Writing – review & editing. **Fernando Hevia**: Formal analysis, Funding acquisition, Writing - original draft, Writing – review & editing.

# Acknlowledgements

J. V. A.-L. would like to thank the Instituto de Corresponsabilidade pela Educação (ICE) – Brazil for his PhD scholarship. F. P. acknowledges the FPI grant PREP2022-000047 from MCIN/AEI/10.13039/501100011033/ and FEDER, UE.

# Funding

This work was supported by Project PID2022-137104NA-I00 funded by MCIN/AEI/10.13039/501100011033/ and by FEDER, UE.